\documentclass{article}
\usepackage{spconf,amsmath,graphicx,hyperref}

\usepackage[utf8]{inputenc}

\usepackage[linesnumbered,ruled,vlined]{algorithm2e}
\usepackage{tabularx}
\SetKwInput{KwIn}{Input}
\SetKwInput{KwOut}{Output}
\usepackage{hyperref}  
\usepackage{booktabs}  
\usepackage{array}     
\usepackage{multirow}  
\usepackage[ruled,vlined]{algorithm2e}
\usepackage{cite}
\usepackage{amsmath,amssymb,amsfonts}
\usepackage{algorithmic}
\usepackage{graphicx}
\usepackage{textcomp}
\usepackage{xcolor}
\def\BibTeX{{\rm B\kern-.05em{\sc i\kern-.025em b}\kern-.08em
    T\kern-.1667em\lower.7ex\hbox{E}\kern-.125emX}}
\title{Bridging the Gap between Training and Inference in LM-based TTS models}
\name{Ruonan Zhang\textsuperscript{1}, 
Lingzhou Mu\textsuperscript{1}, 
Xixin Wu\textsuperscript{2}, and
Kai Zhang\textsuperscript{1,*}
}
\address{\textsuperscript{1}Tsinghua University,
\textsuperscript{2}The Chinese University of Hong Kong}
\begin{document}
%
\maketitle
\begin{abstract}
Recent advancements in text-to-speech (TTS) have shown that language model (LM) based systems offer competitive performance compared to traditional approaches. However, in training, TTS models use ground-truth (GT) tokens as prefixes to predict the next token, while in inference these tokens are not available, a gap between training and inference that is often neglected. In this study, we propose a prompt-guided hybrid training scheme to mitigate exposure bias in popular LM-based TTS systems. Our core idea is to adopt a hybrid training paradigm that combines teacher forcing with free running, thereby introducing self-generated tokens into the training process. This makes the training mode more consistent with inference, reducing the training–inference gap. In addition, we incorporate an EOS prediction mechanism during training to detect incorrect sequence termination and adaptively control the free running process. Experimental results provide a comprehensive evaluation of the impact of exposure bias on LM-based TTS, and demonstrate that our method effectively narrows the training–inference gap, thereby improving the quality of synthesized long-form speech.

\end{abstract}
\section{Introduction}
\label{sec:intro}

Recent years have witnessed rapid advancements in TTS synthesis models, enabling them to generate highly natural and expressive speech.\cite{li2019neural,song2024ella}. These models commonly discretize speech signals into token sequences and adopt autoregressive next - token prediction, leveraging the generative capabilities of pre-trained Large Language Models (LLMs) for sequence modeling \cite{meng2024autoregressive}. Most LM-based TTS models\cite{chen2024enhancing,fu2024asrrl,gao2025emo} suffer from a fundamental limitation, the training - inference gap, commonly known as exposure bias\cite{bengio2015scheduled,zhang2019bridging}. While training relies on teacher forcing with GT tokens as input, inference requires the model to generate tokens autoregressively based on its own previous predictions\cite{chen2024f5,meng2024autoregressive}. This discrepancy between training and inference often leads to cascading prediction errors.

Exposure bias is a common challenge in autoregressive sequence modeling tasks\cite{fang2023understandingbridgingmodalitygap,wang2020exposurebiashallucinationdomain,zhou2025indextts2breakthroughemotionallyexpressive}.
In text generation tasks such as neural machine translation\cite{fang2023understandingbridgingmodalitygap}, exposure bias may lead to semantic drift. To alleviate exposure bias, scheduled sampling \cite{bengio2015scheduled,zhang2019bridging,duckworth2019parallel} is proposed to mitigate the training-inference gap.
However, its impact is magnified in speech synthesis due to the high density of acoustic tokens. 
Unlike text, token-based speech synthesis operates at a much higher temporal resolution, where minor errors are not as easily tolerated. This high resolution typically requires 50–100 discrete tokens per second to represent continuous speech.
A single syllable or phoneme is often mapped to several tokens. Consequently, even minor prediction errors can spread rapidly and negatively impact the listener's experience.
Moreover, speech exhibits additional attributes such as speaking rate and prosody, which are absent in textual representations but play a important role in evaluation in TTS.
This effect amplifies the impact of exposure bias, particularly for long-form speech synthesis.\cite{tian2025preference,li2025styletts}.  

The manifestations of exposure bias are diverse. First, TTS models tend to generate speech at an accelerated speed during long utterance synthesis. The second issue is EOS misprediction, which results in early stopping or repeating parts of the speech. Additionally, the synthesized speech exhibits excessive prosodic flatness relative to ground truth. Thus, training-inference gap is obviously a significant challenge for LM-based TTS models.
\begin{figure*}[h]
  \centering
  \includegraphics[width=1\linewidth,keepaspectratio]{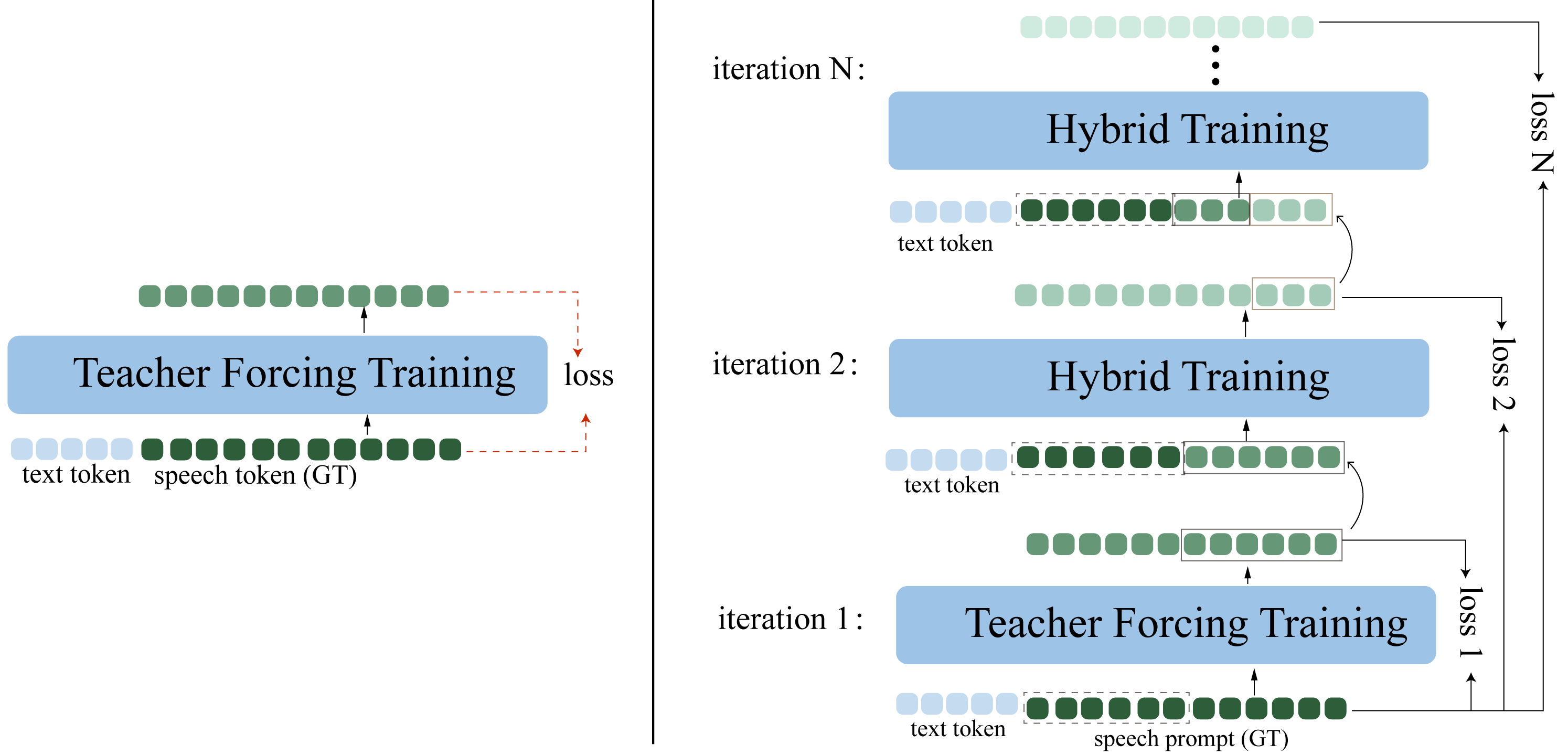}
  \caption{The proposed Hybrid framework.}
  \label{framework}
  
\end{figure*}

To mitigate exposure bias, we propose a prompt-guided hybrid training scheme that gradually transitions from fully guided supervision to self-conditioned generation. As illustrated in Figure~\ref{framework}, our approach operates iteratively within each training step. First, it replaces a portion of the GT input tokens with self-generated tokens in a previous pass. Concurrently, we use a prompt protection strategy to ensure a specific part of the GT tokens is always preserved. This controlled exposure not only maintains training stability but also enhances generalization. We further introduce an adaptive free running strategy guided by End-of-Sequence (EOS) prediction. If the EOS is mispredicted, we end free running early by masking subsequent steps and proceed directly to gradient updates. 
Our experimental results demonstrate that the iterative hybrid training strategy enhances the model's dependency on its token history, compensating for the absence of GT tokens during inference. This approach also narrows the gap between training and inference modes, leading to more stable and coherent sequence modeling. 



\section{proposed approach}
\label{sec:format}
\subsection{Prompt-guided hybrid training scheme}
Figure ~\ref{framework}(right) illustrates our training framework in comparison with standard teacher forcing training (left). We propose a prompt-guided hybrid training scheme that integrates teacher forcing with free running. The basic idea of our training strategy is to iteratively put the self-generated tokens into the LLM being trained to simulate auto-regressive inference process and back propagate both teacher forcing loss and accumulated free running loss in a single training step. To further strengthen its prompt following ability, we randomly replace the starting speech tokens with GT tokens in later iterations. This design leverages the stability of teacher forcing while gradually introducing self-generated prefixes, thereby improving robustness against exposure bias.

Moreover, when prompt protection strategy is enabled and the first few speech tokens of input sequence is replaced by GT tokens, the replaced portion also provides an additional teacher forcing loss to further stabilize training process. As training progresses, we increase the number of replaced tokens to make a smooth shift from initial teacher forcing dominated training to self-generated token dominated training. This fosters a smooth transition, guiding the model toward robust and reliable self-generation.

\subsection{Adaptive free running scheduling via EOS prediction}
\label{sec:EOS}
Earlier studies have found that teacher forcing demonstrates greater stability and faster convergence compared to free running. Therefore, too many iterations of free running may interfere with model convergence. To address this issue, adaptive free running scheduling via EOS prediction is introduced as a compensatory solution. 

We observe that the predicted EOS token serves as a reliable indicator of exposure bias and training stability, allowing dynamic adjustment of free running iterations. When the EOS token is not correctly predicted, it suggests potential exposure bias or model degradation during training. In such cases, the model should increase its reliance on ground-truth supervision to correct its learning trajectory in subsequent steps. Conversely, when EOS token is successfully predicted in several consecutive training steps, the model increases the number of free running iterations to better align the inference procedure. This adaptive mechanism dynamically adjusts the number of free running iterations based on output quality, thereby improving both training efficiency and generation stability.
\vspace{-3mm}
\subsection{Training object}

The overall training objective is influenced by two components: teacher forcing loss, which provides stable supervision using GT tokens, and a weighted sum of free running losses, which improves prediction quality under autoregressive conditions. As depicted in Figure~\ref{framework}(right), the first iteration of our training framework involves computing the teacher forcing loss $\mathcal{L}_{\text{TF}}$, identical to the cross-entropy loss in conventional LLM training. $\mathcal{L}_{\text{TF}}$ can be described as follow:

\begin{equation}
\mathcal{L}_{\text{TF}} = -\sum_{t=1}^{T} \log P(y_{t} \mid y_{i<t}^{\text{gt}}, \mathbf{X})
\label{eq:losstf}
\end{equation}

where, $y_{t}$ denotes the $t_{\text{th}}$ speech token generated from its GT prefixes $y_{i<t}^{\text{gt}}$. $\mathbf{X}$ refers to the input prompt tokens, including SOS token, text embeddings, speaker embedding and other prompt tokens.

During the second to the last iteration, we replace the first $T1$ tokens of the output from the previous iterations with GT and then input this modified sequence into the model. We compute cross entropy loss between model output with GT tokens as follow:
\vspace{-2mm}
\begin{multline} 
\label{eq:lossfr}
\mathcal{L}_{\text{FR}} = - \sum_{t=1}^{T_1} \log P(y_t \mid y_{i<t}^{\text{gt}}, \mathbf{X}) \\ 
- \sum_{t=T_1+1}^{T_2} \log P(y_t \mid y_{i<T_1}^{\text{gt}}, y_{T_1<i<t}^{\text{pred}}, \mathbf{X})
\end{multline}

This free running loss $\mathcal{L}_{\text{FR}}$ is a sum of two terms. The first term refers to the cross entropy between GT and the first $T_1$ speech tokens of model output. These tokens are generated based on GT prefixes $y_{i<t}^{\text{gt}}$ and $X$. The second term refers to the cross entropy between GT and the rest of the speech tokens which is generated based on GT tokens $y_{i<T_1}^{\text{gt}}$, predicted tokens $y_{T_1<i<t}^{\text{pred}}$, and $X$. Our overall training objective can be formulated as follow:

\begin{equation}
\mathcal{L}_{\text{total}} = \mathcal{L}_{\text{TF}} + \sum_{n=1}^N w_n \mathcal{L}_{\text{FR}}^{(n)}
\label{eq:total_loss}
\end{equation}

where \(\mathcal{L}_{\text{FR}}^{(n)}\) represents the free running loss from iteration \(n\), and $w_n$ is its associated weight. We use the weighted average of the loss from each iteration as the final loss to balance the effects across iterations.

\section{Experiments}

\subsection{Implementation details}
The proposed prompt-guided hybrid training scheme is employed to fine-tuning the CosyVoice \cite{du2024cosyvoice} and CosyVoice2\cite{du2024cosyvoice2scalablestreaming} on the LibriSpeech corpus\cite{pratap2020mls}, which contains approximately 40K hours of transcribed speech data. For evaluation, we adopt two test sets, LibriSpeech test-clean and SeedTTS \cite{anastassiou2024seed}, which includes 1,000 utterances sampled from the common voice corpus to reflect more diverse acoustic conditions. 
During training, we adopt the AdamW optimizer with a constant learning rate of $1 \times 10^{-5}$. Speech signals are discretized into token sequences using a vector quantizer with a single codebook of 4096 entries. We apply 10k warm-up steps before introducing the proposed hybrid training strategy.

\subsection{Subjective evaluation}
\label{sec:subj}
As shown in Table~\ref{tab:exp_fig1}, we compare against the baselines IndexTTS\cite{deng2025indexttsindustriallevelcontrollableefficient}, CosyVoice and CosyVoice2. CosyVoice-TF denotes fine-tuning the pretrained CosyVoice model using the standard teacher forcing scheme. CosyVoice2-TF is defined analogously. On CosyVoice2, our prompt-guided hybrid training strategy achieves the best overall performance. It yields superior Word Error Rate (WER) and speaker smilarity performance over baselines on LibriSpeech and relatively smaller improvement on Seed-TTS (where utterances are $<10s$). The relatively small improvement on Seed-TTS dataset may be attributed to the shorter duration of its speech samples, as our method tends to perform better on longer speech samples by mitigating the accumulated error caused by exposure bias. It is also quite intuitive that longer sequence is easier to suffer from accumulated error in the process of autoregressive generation. 

\begin{table}[htbp]
  \centering
  \caption{Comparison results with existing methods. }
  \label{tab:exp_fig1}
\setlength{\tabcolsep}{3.5pt} 
\begin{tabularx}{\columnwidth}{@{}l XX XX @{}} 
    \toprule
   System & \multicolumn{2}{c}{LibriSpeech} & \multicolumn{2}{c}{Seed-TTS}\\
    \cmidrule(lr){2-3} \cmidrule(lr){4-5} 
    {} & WER $\downarrow$ & SIM$\uparrow$ & WER $\downarrow$ & SIM$\uparrow$ \\
    \midrule
    GT & 1.94 & 0.82 & 2.14 & 0.89 \\
    \midrule
    IndexTTS &8.30 & 0.68 & 6.53 & 0.84\\
    CosyVoice - TF & 8.17 & 0.67 & 6.54 & 0.78 \\
    CosyVoice2 - TF & 6.23 & 0.74 & 4.83 & 0.83 \\
    
    \midrule
     CosyVoice - Ours & 5.38 & 0.68 & 6.34 & \textbf{0.85} \\
    CosyVoice2 - Ours & \textbf{4.21} & \textbf{0.78} & \textbf{4.64} & 0.84 \\
    \bottomrule
\end{tabularx}
\setlength{\tabcolsep}{6pt} 

\end{table}

\subsection{Objective evaluation}

\begin{figure}[htbp]
  \centering
\includegraphics[width=1\linewidth,keepaspectratio]{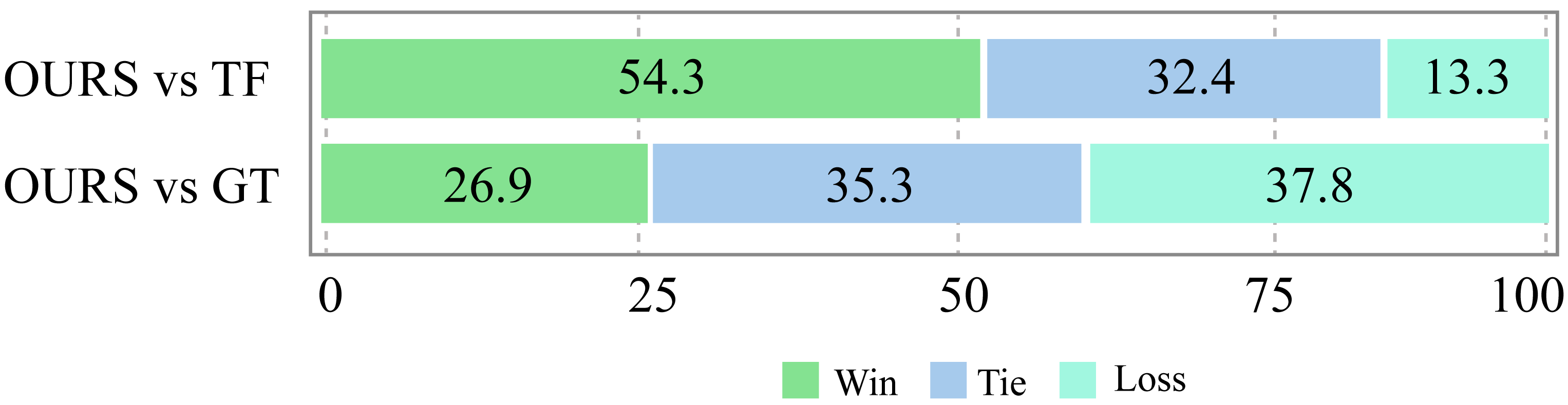}
  \caption{Objective Evaluation}
  \label{fig:ab_results}
\end{figure}

For objective evaluation, we evaluate our method using both mean opinion score (MOS) and A/B preference tests with human listeners. 
Thirty samples are obtained from the LibriSpeech test-clean set, and the MOS results are shown in Figure ~\ref{fig:ab_results}. Our method outperforms teacher forcing training and yield comparable results with GT speech samples.

\subsection{Visualization of exposure bias}
\begin{figure}[h]
  \centering
  \includegraphics[width=1\linewidth,keepaspectratio]{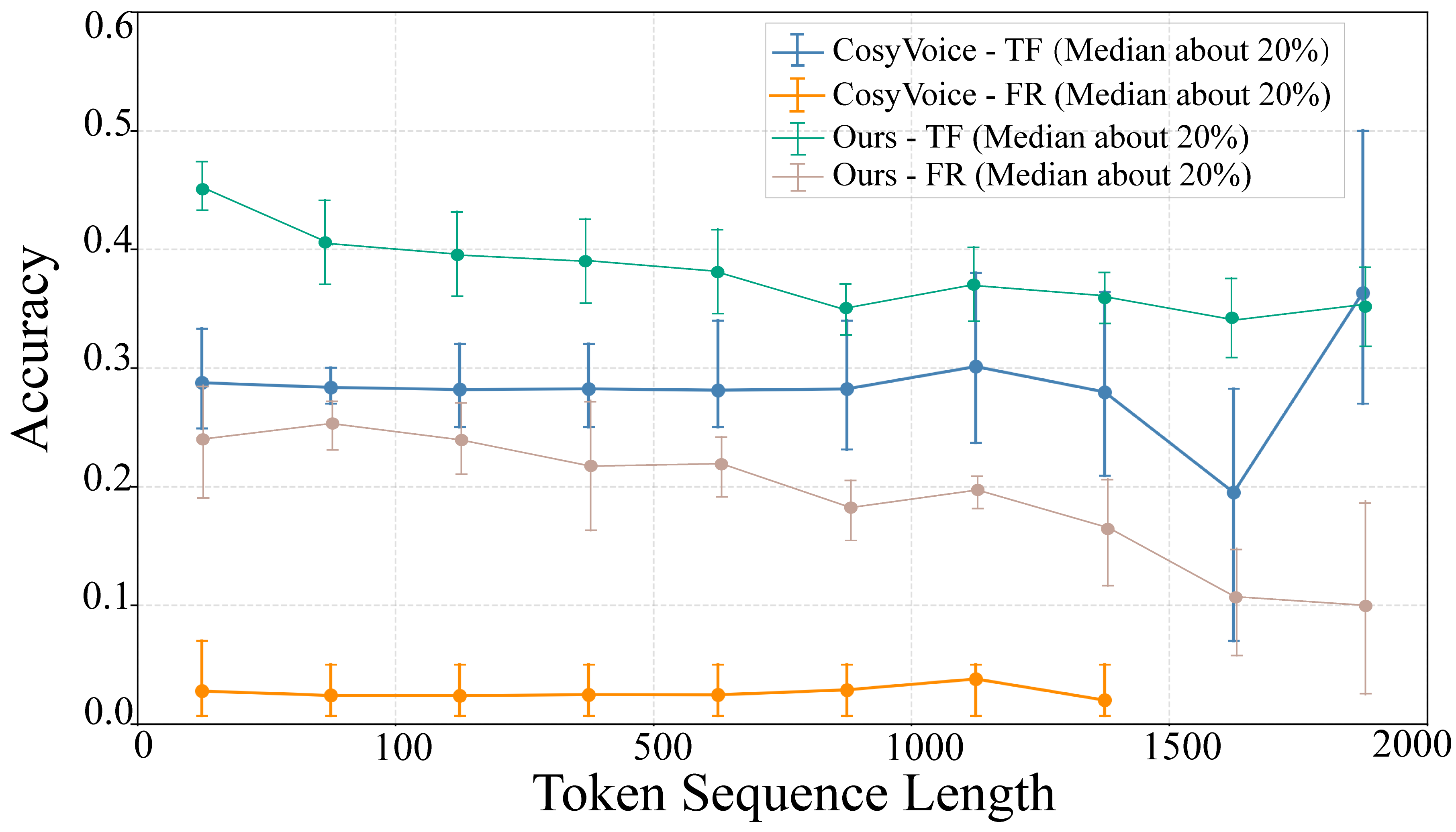}
 
  \caption{Exposure bias illustrated by accuracy gap between teacher forcing and free running.}
  \label{fig:exposurebias}
\end{figure}

To visualize the impact of exposure bias, we compute token-level accuracy by comparing the predicted token index with the ground-truth token index at each position and calculating the proportion of exact matches. 
As shown in Figure~\ref{fig:exposurebias}, the blue line indicates the token prediction accuracy under teacher forcing, where GT tokens are used as context, while the yellow line corresponds to the accuracy when tokens are generated auto-regressively. We point out that speech token accuracy is inherently low, which is determined by the nature of speech token. To be specific, after speech is tokenized, similar sounds can be represented by different speech tokens, making sound-token mapping is not strictly one-to-one.

Despite the low accuracy, the relative difference between teacher forcing and free running inference remains informative for analyzing exposure bias.
As demonstrated in Figure ~\ref{fig:exposurebias}, a clear accuracy gap appears between teacher forcing (blue) and free-running inference (yellow). This discrepancy highlights the presence of exposure bias. In contrast, our hybrid-trained model not only boosts the absolute accuracy in both modes but also significantly narrows the gap between them.

\vspace{-2mm}
\subsection{Validation of EOS-guided scheduling}

\begin{figure}[h]
  \centering
 \includegraphics[width=1\linewidth,keepaspectratio]{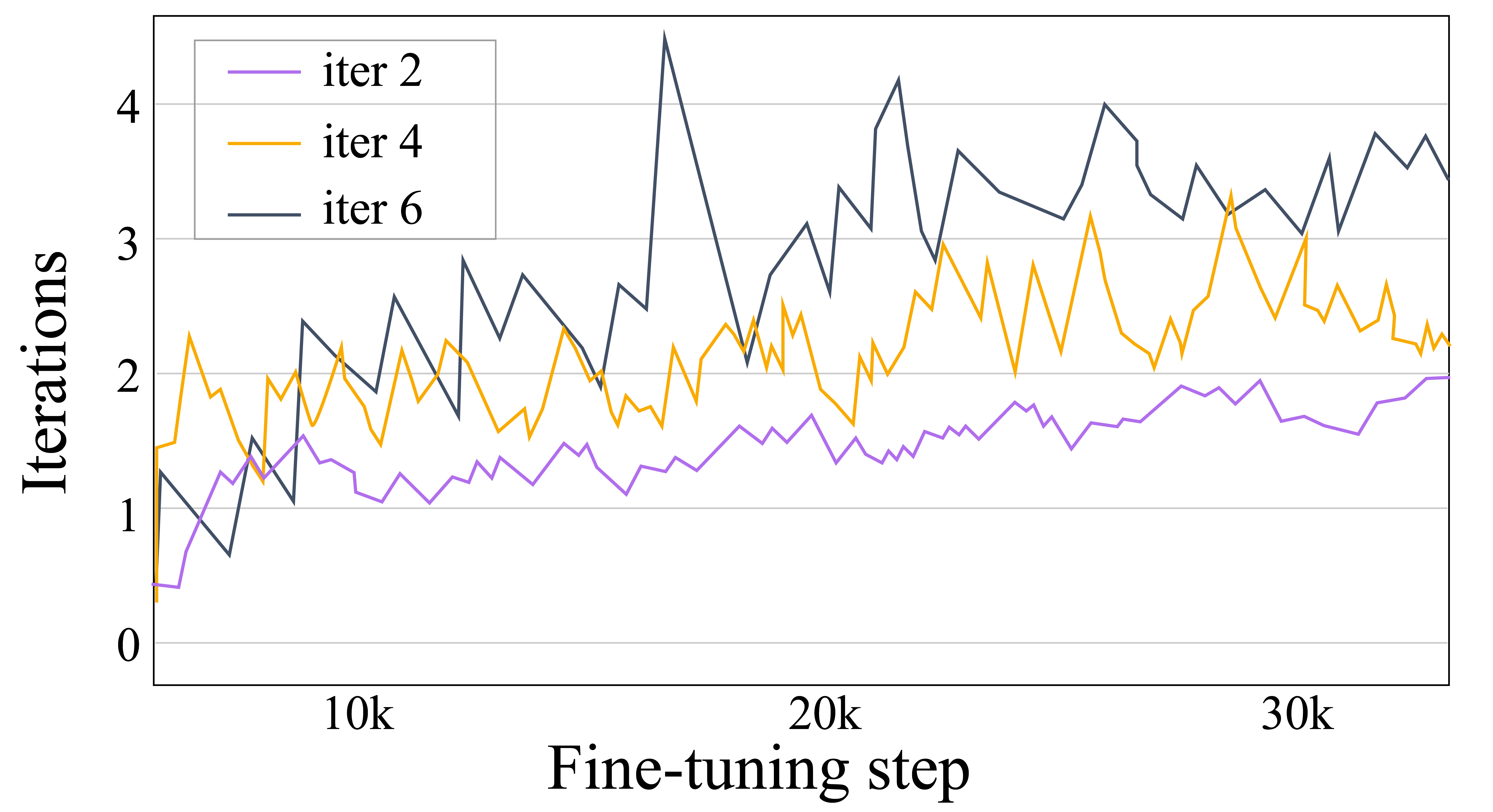}
  \caption{Frequency of early termination vs. iteration number. The rising trend indicates improved EOS error detection and adaptive free running, with increased quality control at the cost of computation.}
  \label{fig:step}
\end{figure}

To validate EOS guided adaptive iteration introduced in Section \ref{sec:EOS}, we track the number of iterations before the model incorrectly predicts a premature EOS token. 
Experiments are conducted under different maximum iteration settings (2, 4, and 6). As shown in Figure ~\ref{fig:step}, during the early phase of hybrid training, premature EOS often occurs within the first few iterations, suggesting that the model is not fully prepared for free running. As training goes on, the occurrence of premature EOS shifts to later steps, suggesting improved robustness under free running conditions. Based on this observation, we increase the number of free running iterations when premature EOS predictions are less frequent. Notably, despite the gradual increase in iterations during training, our method requires only 1.5× the computation of the baseline and converges in fewer steps. 
This training efficiency attributed to the fact that we only do backward pass once in a training step.

\subsection{Ablation Study}

As summarized in Table \ref{tab:exp_fig2}, we conduct an ablation study to validate the effectiveness. The results show that removing either prompt protection or EOS adaptive leads to higher WER and lower speaker similarity. The degradation is more pronounced when Prompt Protection is removed.
\begin{table}[htbp]
  \caption{Ablation Study. w refers to our training framework with all method enabled.}
  \label{tab:exp_fig2}
  \centering
  \begin{tabularx}{\columnwidth}{@{}l XXX @{}}
    \toprule
    \textbf{systems} & WER $\downarrow$ & SIM $\uparrow$ \\
    \midrule
    \textbf{w/o Prompt Protection} & 7.25 & 0.65 \\
    \textbf{w/o EOS Adaptive} & 4.98 &0.72  \\
    \textbf{w} & \textbf{4.21} & \textbf{0.80} \\
    \bottomrule
  \end{tabularx}
\end{table}
\vspace{-3mm}
\section{Conclusion}

We propose a novel training framework for LM-based TTS to align the training process with autoregressive inference. By constructing a hybrid input that mixes GT tokens with the model's self-generated tokens, our method effectively mitigates exposure bias. This strategy effectively simulates the data distribution encountered during inference. The prompt protection mechanism guides the model to gradually shift from supervised guidance to self-sufficient autoregressive generation. Concurrently, it offers a reliable feedback EOS signal for dynamically tng the free-running scheduling strategy.
In our experiments, we first visualize the significant exposure bias between the training and inference stages. Furthermore, the framework alleviates the problem of the model struggling to predict the EOS token during free running generation, a classic symptom of exposure bias. The results demonstrate that our proposed training framework significantly improves speech synthesis quality, especially for long-form speech synthesis.



\vfill\pagebreak


\bibliographystyle{IEEEbib}
\bibliography{refs}

\end{document}